# Deep learning selection of analogues for Mars landing sites in the Qaidam Basin, Qinghai-Tibet Plateau

Fanwei Meng[a,b], Xiaopeng Wang[c,d*], André Antunes[e], Jie Zhao[c], Guoliang Zhou[c], Biqiong Wu[d], Tianqi Hao[b]

[a.] Key Laboratory of Coalbed Methane Resources and Reservoir Formation Process, Ministry of Education, China University of Mining and Technology, Xuzhou, Jiangsu, 221008, China

[b.] School of Resources and Geosciences, China University of Mining and Technology, Xuzhou, Jiangsu, 221116, China

[c.] Yichang Key Laboratory of Robot and Intelligent System, China Three Gorges University, Yichang, 443002, China

[d.] Hubei Key Laboratory of Intelligent Yangtze and Hydroelectric Science, China Yangtze Power Co., Ltd., Yichang, 443133, Hubei

[e] Institute of Science and Engineering, University of Saint Joseph, Macau SAR, Macau, China

* Corresponding author at: Yichang Key Laboratory of Robot and Intelligent System, China Three Gorges University, Yichang, 443002, China

E-mail address: xiaopengwang@ctgu.edu.cn (XP Wang)

ABSTRACT:

Remote sensing observations and Mars rover missions have recorded the presence of beaches, salt lakes, and wind erosion landforms in Martian sediments. All these observations indicate that Mars was hydrated in its early history. There used to be oceans on Mars, but they have now dried up. Therefore, signs of previous life on Mars could be preserved in the evaporites formed during this process. The study of evaporite regions has thus become a priority area for Mars' life exploration. This study proposes a method for training similarity metrics from surface land image data of Earth and Mars, which can be used for recognition or validation applications. The method will be applied in simulating tasks to select Mars landing sites using a selecting small-scale area of the Mars analaogue the evaporite region of Qaidam Basin, Qinghai-Tibet Plateau. This learning process minimizes discriminative loss function, which makes the similarity measure smaller for images from the same location and larger for images from different locations. This study selected a Convolutional Neural Networks (CNN) based model, which has been trained to explain various changes in image appearance and identify different landforms in Mars. By identifying different landforms, priority landing sites on Mars can be selected.

*Keywords*: Remote sensing; evaporite; Convolutional Neural Networks; Qaidam Basin; landing sites; analogue

## 1. Introduction

Based on its geological time scale, Mars is divided into the Noachian era (4 billion to 3.7 billion years ago), Hesperian era (3.7-3 billion years ago), the Amazonian era (3 billion years ago to present). A large amount of geological data supports that there was a rich hydrosphere on Mars during the Noachian period, with a scale equivalent to 100-1500 m global equivalent layer (GEL) (Di Achille and Hynek, 2010). This relatively large volume of liquid water on the surface of Mars gradually disappeared over 3 billion years ago (Scheller et al., 2021), with the escape of water vapor and gradual increase in salinity, ending with the formation of evaporites (Knauth, 2005; Murchie et al., 2009). Indeed, satellite photos and preliminary spectrum analysis seems to confirm their existence in several locations on Mars (Tosca and McLennan, 2006; Osterloo et al., 2008; Edwards et al., 2009; Ehlmann and Edwards, 2014). Furthermore, it cannot be ruled out that underground brines are still present in Mars (Bridges and Schwenzer, 2012).

The setting and evolution on our planet were different but can nonetheless provide some useful insights. Earth originated 4.5 billion years ago, with a very small original land area, which gradually increased with time (Burnham and Berry, 2017). NaCl (halite) in seawater was gradually deposited and preserved in sedimentary basins on various continents, so the salinity of the original ocean gradually decreased (Figure 1). These deposits are quite extensive: estimates suggest that, if all the halite preserved in these basins were dissolved in our current oceans, its salinity would increase by 1.6-2 times or more (Holland, 1978; Knauth, 2005; Meng et al., 2011). The higher salinity of this primordial ocean has led some authors to suggest that the oldest life forms on Earth might have been salt tolerant or halophilic microorganisms (Knauth, 2005).

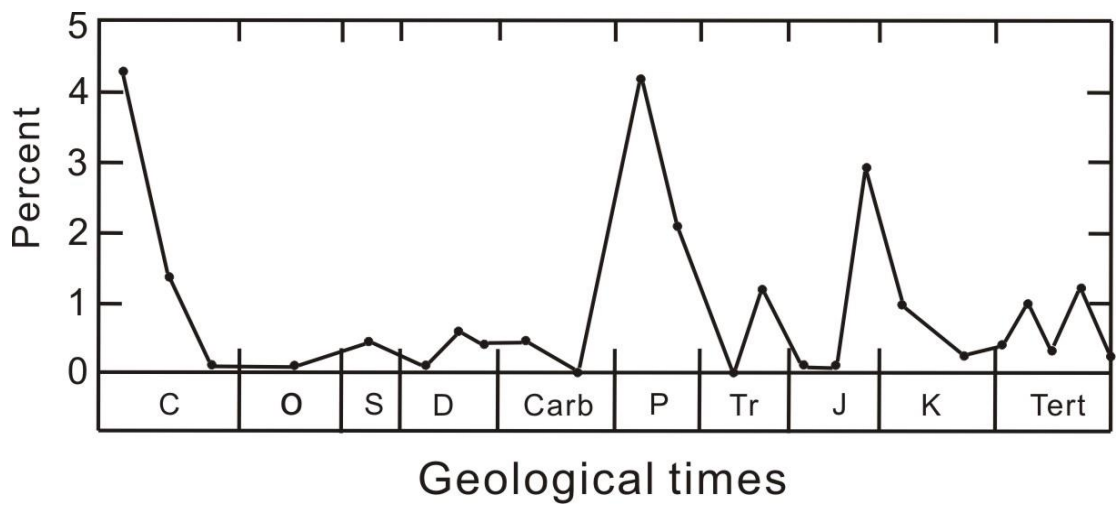

**Fig. 1** Changes in abundance of evaporites through Phanerozoic time (revised from Ronov et al., 1980).

When looking at evaporation of seawater on Earth and the process of evaporite formation in modern inland salt lakes and marine lagoons, we see a typical precipitation sequence. First, sulfates (e.g., gypsum and anhydrite deposits) are deposited, followed by halite and eventually by potassium magnesium salt minerals (Warren, 2010). Halite is the most important and most abundant mineral after the complete drying up of the evaporite basin (Figure 2). From a biological perspective, the increase in salinity and even salt crystal precipitation is not contrary to life. Indeed, microbes have been shown to be incredibly resilient, thriving under high salinity conditions and even surviving over geological timescales when trapped inside salt crystals (Antunes et al., 2017; Stan-Lotter and Fendrihan, 2015). Such halophilic microbes, particularly members of the archaeal class *Halobacteria* are considered key model organisms for Astrobiology (Baxter et al., 2007; DasSarma, 2006; Wu et al., 2022).

Given the links between ocean drying, evaporite formation, and long-term preservation of microbial life in such deposits, this is an increasingly important topic regarding the search for Life on Mars. Accrodingly, when searching for prime locations to collect samples or look for signs of past or present Life on Mars, priority should be given to areas with evaporites. Locations on Earth with similar evaporite landforms to these observed on Mars provide ideal locations for comparative research on the distribution characteristics and formation mechanisms of Martian related landforms. Based on this, Martian analog studies are frequently conducted and look at geomorphological features（Zheng et al., 2013）, as well as on microorganisms present in evaporite minerals such as gypsum and halite have been conducted by several authors (e.g., Kong et al., 2010; Conner and Benison, 2013; Zheng et al., 2013; Benison and Karmanocky III, 2014) (Figure 3).

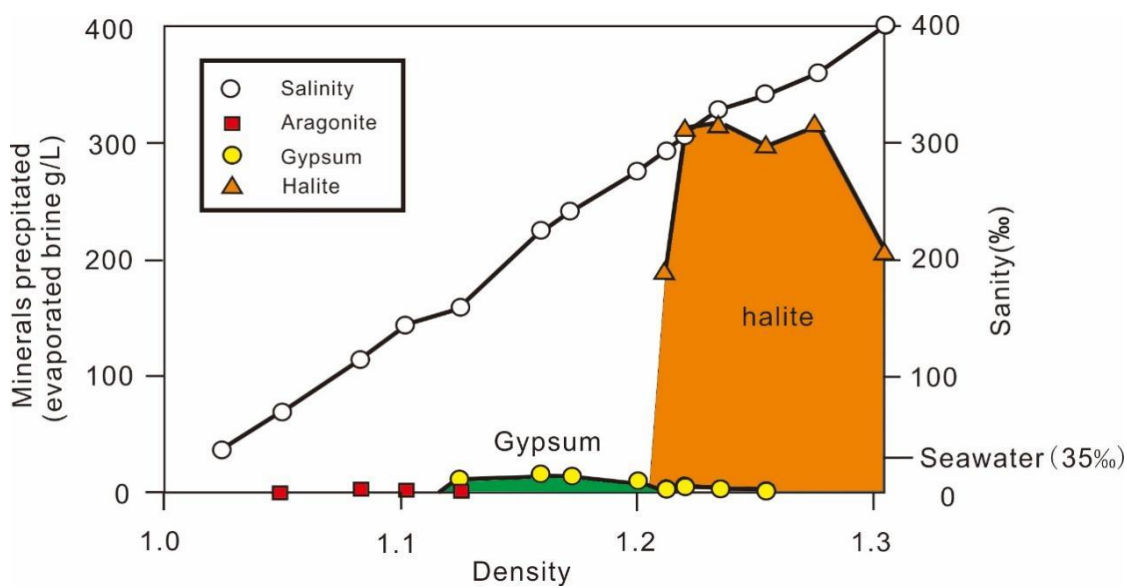

**Fig. 2** Mineral precipitation sequence in concentrating marine brines before bittern salts (revised from Briggs 1958).

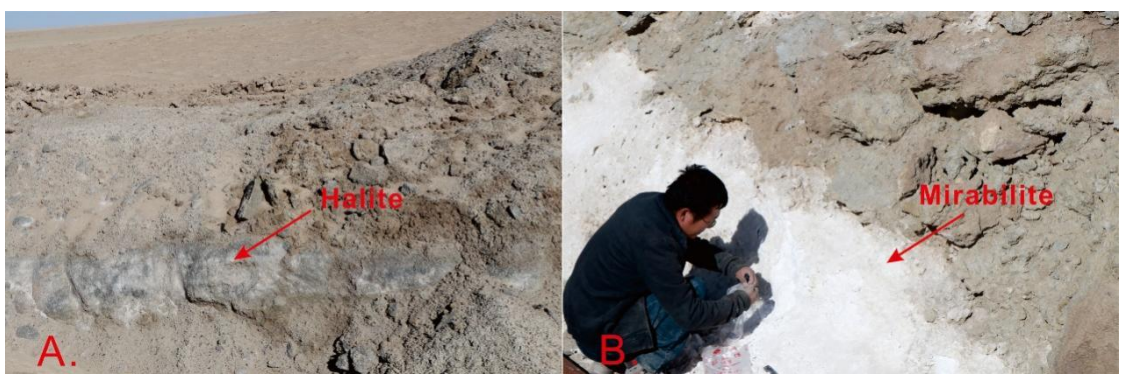

**Fig. 3** Evaporites in Qadam Basin (A. Halite; B. Mirabilite).

Together with ice, halite is acknowledged as one of the best materials for long-term preservation of microorganisms (Meng et al., 2018a,b) and/or their biosignatures. Part of this success is a direct result from the existence of abundant fluid inclusions, captured by lattice defects in halite during the deposition process. There are two types of fluid inclusions in halite: primary fluid inclusions and secondary fluid inclusions. The former are generated during the evaporation and sedimentation stages of brine, typically exhibiting distinct alternating growth bands of fluid inclusions that can be preserved in halite from hundreds of millions of years ago (e.g Kovalevich et al., 1998; Meng et al., 2013) (Figure 4). Secondary fluid inclusions in halite can form at any time after the sedimentary period (Kovalevich et al., 1998). During the formation of these primary fluid inclusions, ancient minerals, microorganisms, biomolecule, and gases can be captured and preserved (Meng et al., 2015) （Fig. 4）.

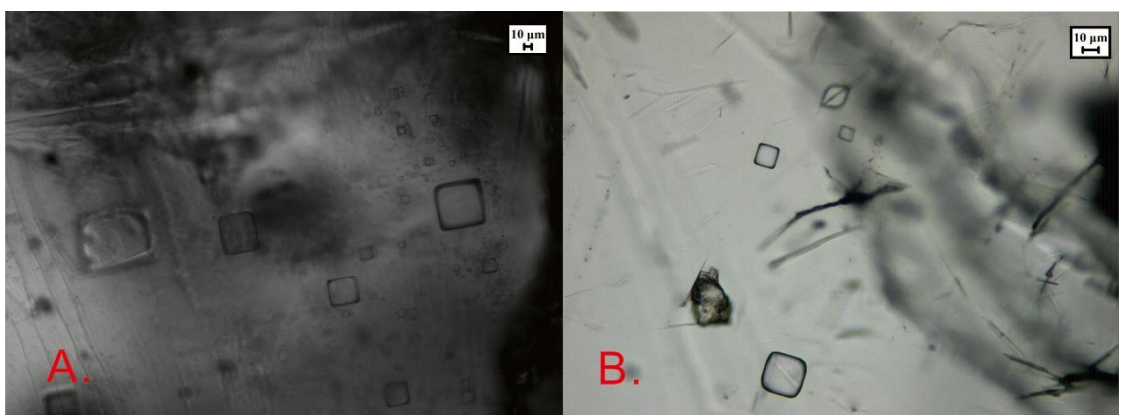

**Fig. 4** Primary Fluid inclusions observed in halite from Qaidam Basin.

The Qaidam Basin (Qinghai-Tibet Plateau) covers an area of 34,900 km$^2$ and is located at an altitude of 2500-3000 meters. The basin's setting makes it the highest average elevation non-polar desert region in the world and exposes it to strong ultraviolet radiation. The Qaidam Basin, as an area with widely distributed evaporites, is considered a key Mars analogue site (Mayer et al., 2009；Zheng et al., 2013; Kong et al., 2013, 2014). The wind-erosion landforms of Qaidam Basin are very diverse, as are its salt lakes and evaporite-linked geomorpholigcal features, and are comparable to the dry and cold conditions of Mars.

In the past 20 years, the rapid development of computer and information technologies have led to remarkable achievements. The booming field of "Artificial Intelligence" and the trend of 'machines replacing human labor' is now spreading to various fields. Artificial intelligence relies on massive data resources and through targeted training, it can enable computers to accurately simulate various human perception and decision-making processes, thereby replacing humans to complete some related work tasks (Li et al., 2018; Liu et al., 2019; Chu et al., 2020; Saleh et al., 2021). One of these fields- Computer vision- relies on the upload information from the external world into computers in the form of images, followed by extraction of valid information, and thus enabling computer-assisted tasks such as detection, recognition, classification, and tracking of targets.

From the human sight perspective, determining target feature similarity is a relatively vague problem. It is very difficult to directly obtain a quantified value to reflect the target features, and it is easily disturbed by external and internal distractions. This is very different when seen from the perspective of a computer. The input image information obtained by the computer is essentially a series of digitized pixel values. The computer must analyze the statistical rules of the target in various dimensions to extract the target area features from a pixel table composed of numbers, thereby obtaining a quantified target feature similarity to reflect the matching problem of target features. Furthermore, the working state of a computer is not affected by distractions.

From this perspective, using computers to carry out target feature similarity analysis has significant advantages. Although scholars have been researching the field of target feature matching for many years, with the emergence of a plethora of target-matching algorithm frameworks based on various theories, several challenges in this field remain unaddressed. Indeed, many uncertainty factors affect the judgment of target feature similarity, such as changes in target surface illumination (Li et al., 2018; Liu et al., 2019), size in the camera's field of view (Ning et al., 2012), appearance/shape (Hui et al., 2017; Chu et al., 2020), target obstruction (Wang et al., 2018; Saleh et al., 2021), and overly complex background information causing interference. Any one of these factors could potentially lead to failure in target matching (Yan et al., 2015).

Applied to the setting of the current paper, this computer-based approach can help to select areas on Earth that are most similar to Martian terrains for simulating Mars landing operations, testing equipment, as well as for selecting priority future landing areas for sample collection and/or detecting signs of past or present life (Liang and Hu, 2015; Wang et al., 2016; Nam and Han, 2016).

Given the previously mentioned challenges, it is of great scientific value to explore robust target similarity algorithms that can better adapt to complex target matching environments for research on Mars-Earth surface geomorphology matching.

**2. Geological Settings**

The western part of the Qaidam Basin sank on the basis of the Mesozoic fault basin, transitioning from a small freshwater lake to a large depression lake. The climate of the Paleocene was dry and resulted in the input of red coarse debris deposits. In the early Oligocene, due to the dry climate and the uplifting of the Altun and Kunlun Mountains, the western part of the basin further subsided, resulting in the earliest salt deposition in the Shizigou area. In the late Pliocene, the uplifting of the Dafengshan-Huangshi structure hindered the supply of fresh water, resulting in a more arid climate and the formation of and precipitation of several salt minerals. From the end of the Pliocene to the early Pleistocene, the salt lake's water was concentrated, and the range of salt deposition expanded. Glauberite, bloedite and polyhalite began to precipitate in the Dalangtan dry salt lake area. In the middle Pleistocene, the structural segmentation of the study area was initially formed, and salt deposits expanded to various depressions, and continued to increase in their extension. In the mid Pleistocene, Dalangtan dry salt lake was isolated, resulting in a smaller lake area and a dry salt beach environment. In the Upper Pleistocene, soluble sulfate, bloedite, and epsomite deposits were formed. 30,000 years ago, tectonic movement continued to sink the basin, and brine converged towards the depression. Soluble potassium and magnesium components converged in seasonal waterlogged depressions, forming chloride type potassium and magnesium salt deposits (Yin et al., 2008a,b) (Fig. 3).

The sedimentary landforms of evaporites on the surface of Mars can be compared with the dry salt lakes areas of the Qaidam Basin, Qinghai-Tibet Plateau, such as Dalangtan dry salt lake (Zheng et al., 2013; Anglés and Li, 2017; Xiao et al., 2017). Indeed, the western Qaidam Basin, (and particularly the Dalangtan dry salt lake area), contains various Martian-like landforms, the most important of which are desert landscapes (such as sand dunes, Yardang, polygonal structures), dry valleys, and salt lakes generated by wind and evaporation.

Polygonal surface structures are a common surface phenomenon of evaporative halite crust in the Qaidam Basin, and this obvious sedimentary structure has also been found on Mars（Zheng et al., 2013; Anglés and Li, 2017; Xiao et al., 2017）. On Earth, the polygonal structure in the salt crust is explained as the growth of salt crystals in the salt and mud layers beneath the dry lake surface during the arid stage of the salt lake sedimentary cycle, resulting in the salt crust splitting into polygons (Christiansen, 1963). Large polygonal structures in salt crusts typically appear on the outer edge of dry salt lakes, while smaller polygonal structures typically appear inside dry salt lakes. The large polygonal structure at the foot of the southern wing of the Dalangtan hill ranges in length from 20 to 150 meters on each side, while the polygonal structure at the bottom of large volcanic craters on Mars has an average size of 120 meters (El Maary et al., 2010), and its characteristics and size are similar to those of the Dalangtan dry salt lake (Fig. 5).

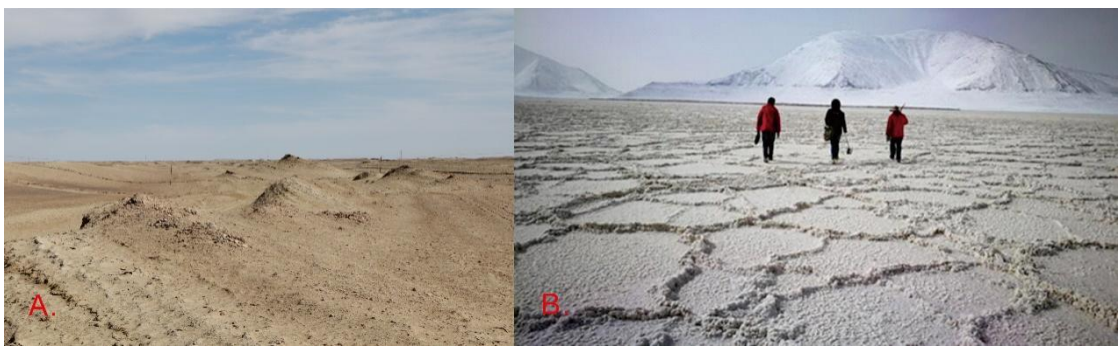

**Fig. 5** Examples of typical geomorphic features from Qaidam, similar to those found on Mars: A. Yardang; B. Large polygonal structures in salt crusts

**Experimental Methods**

In the process of finding the most similar matching Earth terrain image for a Martian terrain image, it is necessary to map the high-dimensional input points of the terrain image to a low-dimensional manifold, to map similar input objects to neighboring points on the manifold. Common data dimensionality reduction methods often fail to generate a function from the input to the manifold that can be applied to new points with unknown relationships to the training points. Secondly, many methods presuppose the existence of meaningful (and computable) distance metrics in the input space. To address this issue, this study employs a method called Dimensionality Reduction by Learning an Invariant Mapping (DrLIM), which is used to learn a globally coherent nonlinear function that maps data evenly to the output manifold. Learning only requires the neighborhood relationships between training samples, which can come from prior knowledge or manual labeling, and are independent of any distance metric. This method can learn a function that is invariant to complex nonlinear transformations of the input, such as changes in illumination and geometric distortions (Hadsell et al., 2006).

To map the original images to points in a low-dimensional space, a network needs to be trained to learn a similarity metric. This study adopts a Siamese network learning architecture, which includes two identical convolutional networks that share the same weight parameter values (Bromley et al., 1993). Convolutional networks are trainable nonlinear learning machines that operate at the pixel level and learn low-level features and high-level representations in an integrated manner (Bell and Bala, 2015). They accept end-to-end training, mapping images to outputs. Due to shared weights and multi-layer structure, they can learn optimal translation-invariant local feature detectors while maintaining invariance to geometric distortions of the input images (Ning et al., 2012; He et al., 2015).

This study uses the high-dimensional data nonlinear dimensionality reduction mapping of convolutional neural networks to carry out regional geomorphological feature similarity analysis of the Qaidam Basin and compares it with Martian terrain. That is, it uses convolutional neural networks to extract low-dimensional feature vectors from the original images and perform comparative analysis, determining whether two targets come from the same geographical area, and thus selecting the geographical location most similar to the target terrain. The algorithm model in this study uses a Siamese network model architecture (Chopra et al., 2005; Zagoruyko and Komodakis, 2015), and analyzes the similarity of the two terrain maps by calculating the Euclidean distance of the low-dimensional feature vectors extracted from the two terrain images. When there is a new sample match test, there is no need to retest the model, just extract the feature vector of the new sample, calculate its Euclidean distance with other sample feature vectors, and then perform terrain matching for the new terrain sample.

## Implementation Methods and Fundamental Theories

We study the method of selecting small-scale Martian land simulation sites in the Qaidam Basin (Fig. 6). Firstly, surface image data of Earth and Mars are collected to build a dataset, which is divided into training and testing sets. A convolutional neural network is established trained and tested to obtain network weight parameters. Then, regions similar to the Martian surface are selected from the Earth surface dataset, and a large number of target images are obtained and their latitudes and longitudes are marked by dividing the image area through the multi-grid method. Secondly, the Martian surface and the divided target images are input into the constructed Siamese network model for comparison all once. Finally, the best matching location is analyzed based on the similarity calculated by the model.

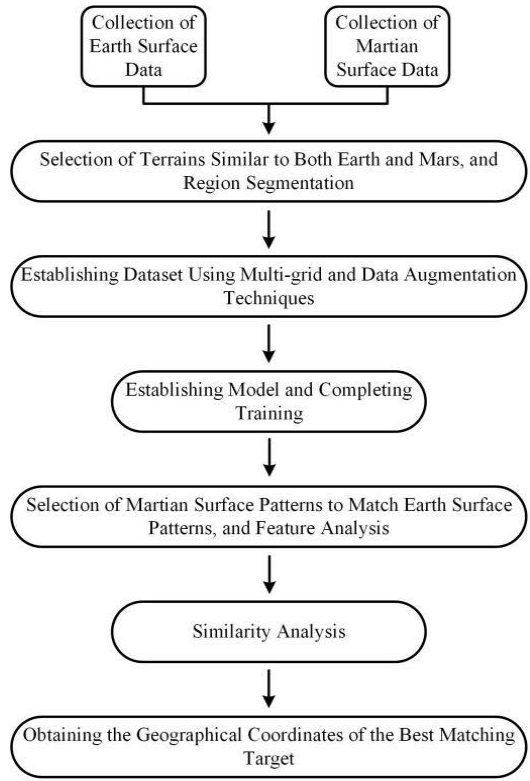

**Fig.6** Process of Selecting Small-scale Martian Land Simulation Sites in the Qaidam Basin.

### 4.1 Convolutional Neural Network Mapping Low-Dimensional Data Features

The experimental data mainly comes from the surface image data of Mars and a small range of the Qaidam Basin, using a convolutional network as a function to map the original image to a low-dimensional manifold. Convolutional neural networks are neural networks that incorporate convolution operations, which can solve problems such as too many parameters in neural networks and too many connections between network layers, maximizing the utilization of network training parameters. The convolutional neural network in this study consists of convolutional layers, nonlinear activation layers, batch normalization (BN) layers, and fully connected layers, etc (LeCun et al., 2015). Among them, the convolutional layer is used for feature extraction. The convolutional layers at the front are responsible for extracting low-level features, such as color, edges, shapes, and the topological structure of the image, etc. Through more convolutional layers at the back, high-dimensional deep features of the image can be extracted. After each convolutional layer, there are ReLU (Rectified Linear Unit) activation functions and BN layers, which are used to accelerate the convergence of the stochastic gradient descent algorithm and prevent overfitting to a certain extent during network training（Fig.7）. The fully connected layer can nonlinearly reorganize the final high-level low-dimensional features (Press et al., 1990).

### 4.2 Similarity Analysis of Low-Dimensional Image Features in Siamese Network Model

The two branches of the Siamese network used in this study have the same weight mapping function (convolutional network), and their feature extraction structure is the same. At the top layer of the network, the output layer is composed of linear fully connected, and nonlinear ReLU activation functions. The output layer contains three fully connected layers, outputting a five-dimensional feature vector to represent the feature information in the input terrain image, and then obtaining its similarity based on the Euclidean distance calculated between the feature information (Chopra et al., 2005).

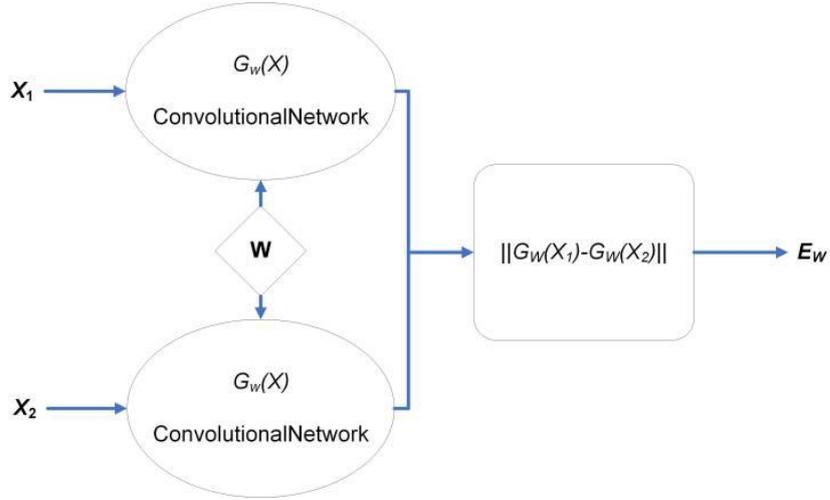

**Fig.7** Structure of the Siamese Network Model.

The learning architecture of the Siamese network model is shown in Fig 7. $X_1$ and $X_2$ are a pair of images input to the model, and Y is the binary label of the image pair. When images $X_1$ and $X_2$ belong to the same target, then Y=0, otherwise Y=1. W is the shared weight parameter obtained by training the convolutional network, and $G_W(X_1)$ and $G_W(X_2)$ are two points generated by mapping $X_1$ and $X_2$ in the low-dimensional space. The parameterized distance function $D_W(X_1, X_2)$ to be learned between $X_1$ and $X_2$ is defined as the Euclidean distance between the outputs of $G_W$, measuring the similarity between $X_1$ and $X_2$, which can be defined as:

$$(1)$$

**4.3** The Contrastive Loss Function

The loss function of the network model uses a contrastive loss function to train the network model. The branch sub-networks of the Siamese network model extract features from terrain image data. The extracted terrain image data is used by this loss function to reduce the distance between similar samples and increase the distance between dissimilar samples, achieving the purpose of network training and improving the network model's ability to distinguish terrain data. The input image data is represented by $(X_1, X_2)$, and the label is represented by Y. If $X_1$ and $X_2$ are samples of the same class, then the value of the label Y is 1, otherwise it is 0. The Euclidean distance between sample labels $D_W(X_1, X_2)$ is simplified as $D_W$. The general form of the contrastive loss function is:

$$(2)$$
$$(3)$$

Where, $(Y, X_1, X_2)^i$ is the i-th labeled sample pair, $L_S$ and $L_D$ are the partial loss functions of a pair of similar samples and dissimilar samples respectively, and P is the number of training pairs.

The design of $L_S$ and $L_D$ needs to ensure that when generating the weight parameter W, minimizing L will be accompanied by a lower $D_W$ value for similar pairs and a higher $D_W$ value for different pairs. That is, a more accurate loss function is:

$$(4)$$

Where, m>0 represents the distance threshold, and m defines the radius around $G_W(X)$. Only when the distance of different pairs is within this radius will they have an impact on the loss function.

### 4.4 Parameter Training

Firstly, for each input sample $X_i$, use prior knowledge or manual labeling to find the set of similar and dissimilar pairs of samples. Add label Y=0 for similar pairs and label Y=1 for dissimilar pairs to form the labeled training set. Then, train the Siamese network model. For each pair in the training set, perform the following calculations:

(1) If Y=0, update W to decrease ;

(2) If Y=1, update W to increase ;

By minimizing the above loss function to increase and decrease the Euclidean distance in the output space, repeat until the loss converges.

### Experimental Verification and Analysis

The experimental environment of the algorithm model proposed in this experiment is shown in Table 1. The operating system is Windows 11, the CPU is Intel Core i7-8750H, the memory is 16 GB, the GPU is NVIDIA GeForce GTX1060, the deep learning framework is Pytorch1.9.0, and the programming language is Python3.9.6. The main extension libraries involved in the model include Matplotlib, torchvision, numpy, etc (Table 1).

Table.1 Experimental environment.

| Types | Details |
| --- | --- |
| Hardware environment | CPU : Intel Core i7-8750H @ 2.20 GHz<br>GPU : NVIDIA GeForce GTX1060<br>RAM : 16GB |
| Software environment | Windows 11<br>Pytorch 1.9.0 |
| Programing language | Python 3.9.6 |

### 5.1 Dataset

This experiment selected images from four representative geomorphic sub-regions within the dataset of the Qaidam Basin and matched them with the central regions of four areas on the surface of Mars. The four Martian terrains mainly include: A. Long-ridge yardangs; B. Linear dunes on Mars; C. Barchans dune chains; D. Polygonal structures. Images of the Martian surface were collected from publicly available articles and databases (Anglés and Li, 2017; Xiao et al., 2017) (Fig 8), while images of the Earth's surface were obtained from Google Earth (Fig 9). Four typical geomorphic features on the Martian surface were selected as research areas, and four similar surface areas were identified in the Qaidam Basin. Using a multi-grid method, the Qaidam Basin and four similar areas on Mars were divided. Each area image was divided into 10 sub-area images and their latitudes and longitudes were marked to establish a dataset (Fig 10).

The images in the dataset were augmented by a factor of 10 through rotation, blurring, scaling, and grayscaling. All data were normalized to ensure that the output image size was 100×100. One image sample was selected from each of the four research areas in the Qaidam Basin and paired with all images from the corresponding research area on Mars, marked as similar. It was also paired with all images from different research areas on Mars, marked as dissimilar. The labeled dataset was randomly divided into training and testing sets at a ratio of 9:1.

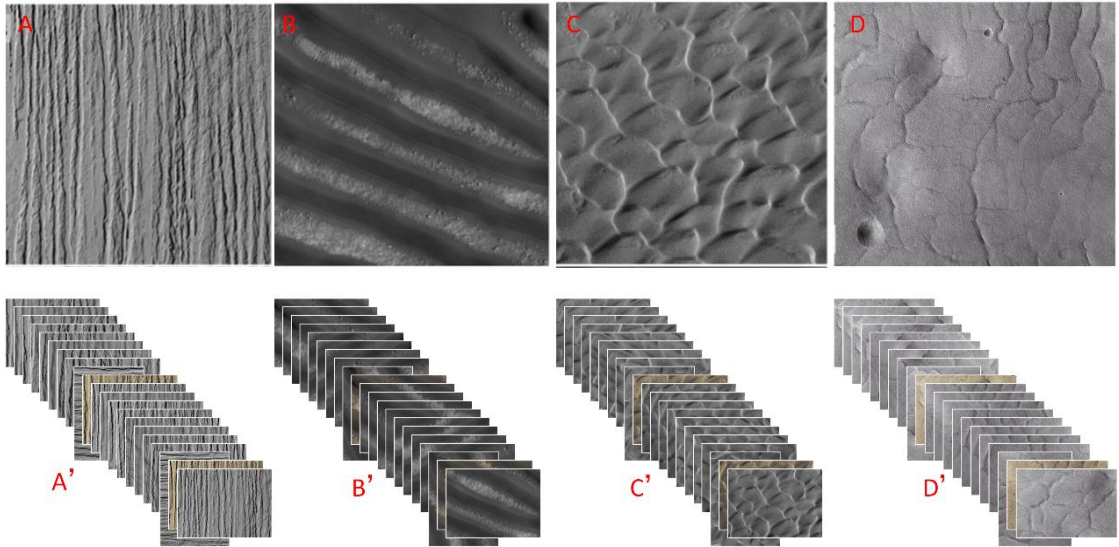

**Fig.8** Selection of the central region on the Martian surface and establishment of the dataset.

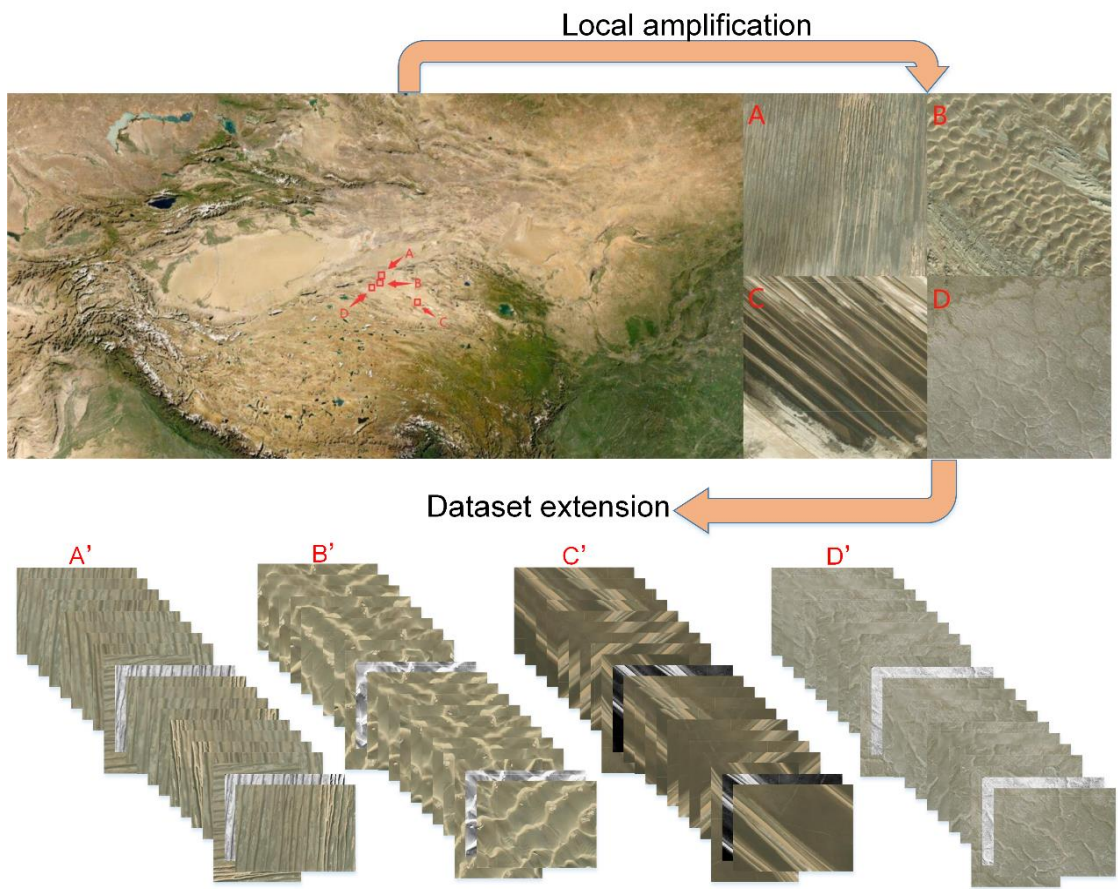

**Fig.9** Selection of representative regions in the Qaidam Basin and establishment of the dataset.

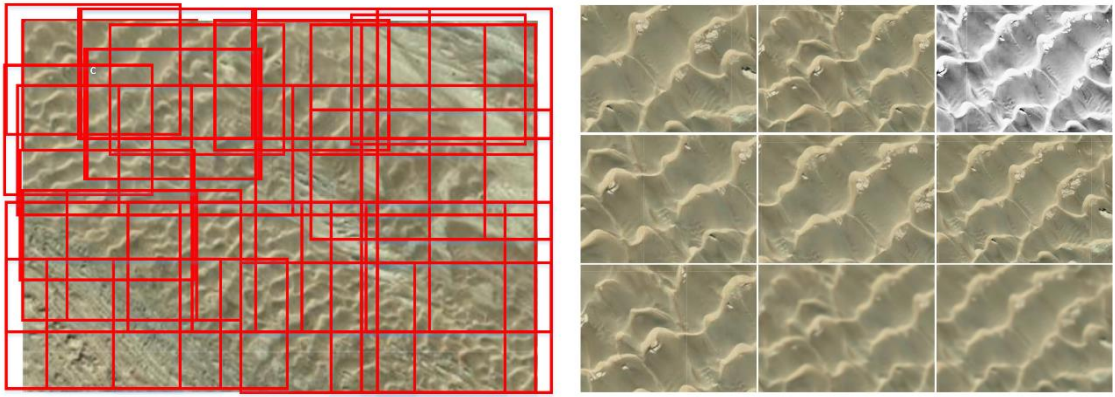

**Fig.10** Selection of local regions and data augmentation.

**5.2** Establishment and training of the Siamese network model

The structure of the convolutional network, which serves as a geomorphic image feature weight mapping function, is shown in Fig.11 (a). This module contains three convolutional layers and two fully connected layers. Considering that small convolution kernels have more non-linearity than large ones (Simonyan and Zisserman, 2014), all convolutional layers use 3×3 kernels, and each fully connected layer is followed by a ReLU activation function. Two identical convolutional networks are constructed to establish the Siamese network model, as shown in Fig.11(b). The model inputs the feature vectors output by the convolutional network into the similarity analysis module, which performs similarity analysis by calculating the Euclidean distance between feature vectors.

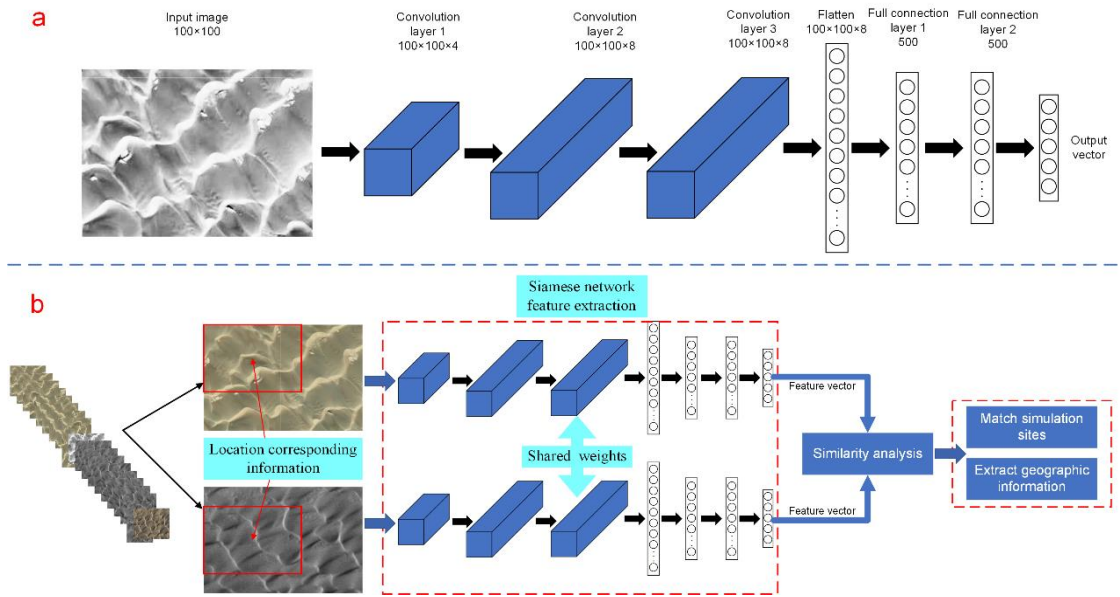

**Fig.11** (a) Architecture of the GW function (Convolutional Network) for learning to map low-dimensional data features; (b) Image similarity analysis framework based on the Siamese network model.

The Siamese network model was trained using the training set. The Adam optimizer was used to optimize the training parameters. The initial learning rate was set to 0.001, and the weight decay value lambda was set to 0.0005. The minimum number of samples for updating parameters in each iteration was 64, and the entire dataset was trained iteratively 300 times. During training, the input to the convolutional network consisted on image data of size 100×100. Considering that the input is two-dimensional data, the convolutional network uses two-dimensional convolution operations. The first, second, and third convolutional layers (C1, C2, and C3) use 4, 8, and 8 convolution kernels respectively, with a stride of 1. At the same time, to ensure that the depth of the output feature map remains unchanged, padding is used around the convolutional layer, resulting in 4, 8, and 8 feature maps of size 100×100 respectively. After completing the third convolutional layer (C3), eight feature maps of size 100×100 are obtained. The two-dimensional features are then unfolded into a one-dimensional vector and input into two fully connected layers for nonlinear mapping. Finally, a five-dimensional feature vector is output as the feature value of the input geomorphic image through a fully connected mapping. The feature values of the input sample pairs are sent to the similarity analysis module, where the similarity value of the input sample pairs is calculated using the Euclidean distance method. During each training iteration, the contrast loss function is used to calculate the loss, and then backpropagation is used to calculate the gradient. After iterating over the entire dataset, the final loss value is 0.0088, and the accuracy of the model is verified on the test set to be 99.82%.

**5.3** Selection of Mars analog sites in the Qaidam Basin

Based on the constructed Siamese network model, an image of the terrestrial surface and an image of Mars are input into the model. The convolutional neural network model is used to extract feature vectors from the input images, and their Euclidean distance is calculated to determine the similarity between the images of Qaidam Basin and the surface of Mars. Four representative geomorphic sub-regions in the Qaidam Basin dataset are selected and matched with the central regions of four areas on Mars.

The locations of the four Martian regions are respectively A'(10.461°S, 176.446°E), B'(42.660°S, 38.023°E), C'(9.12°N, 67.29°E), and D'(341.61°E, 50.70°N). The matched images are input into the Siamese network model to obtain the similarity (or difference) between each small region in the Qaidam Basin dataset and the central region of the four Martian exploration points. Through analysis, we obtained the minimum differences between the four regions of the Qaidam Basin and the four target regions on Mars, as shown in Fig.12. The central location of region A in the Qaidam Basin is (38.493°N, 92.300°E), and the minimum differences with regions A', B', C', and D' on Mars are 0.11, 0.59, 0.72, and 1.04 respectively. The central location of region B in the Qaidam Basin is (37.010°N, 93.991°E), and the minimum differences with regions A', B', C', and D' on Mars are 0.96, 0.26, 1.26, and 0.96 respectively. The central location of region C in the Qaidam Basin is (38.215°N, 91.382°E), and the minimum differences with regions A', B', C', and D' on Mars are 1.28, 0.82, 0.38, and 1.02 respectively. The central location of region D in the Qaidam Basin is (38.395°N, 91.549°E), and the minimum differences with regions A', B', C', and D' on Mars are 0.77, 0.85, 0.76, and 0.05 respectively. We found that the matches A-A', B-B', C-C', and D-D' have the smallest differences of: 0.11, 0.26, 0.38, and 0.05 respectively. Their differences are all less than 0.5, indicating a very high similarity, and can be used as reference objects for Martian analog site (Fig. 12).

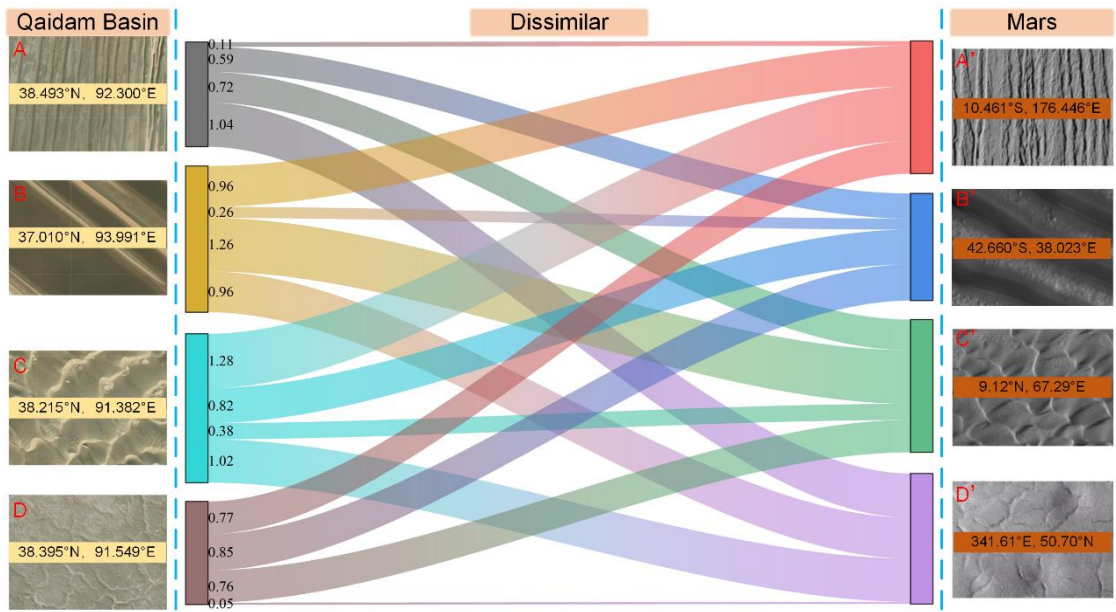

**Fig.12** Matching results between the four regions of the Qaidam Basin and the target regions of Mars.

In the above studies, we confirmed that A-A', B-B', C-C', D-D' have regional similarities. In order to further study the similarity between the simulated points in Qaidam Basin and the Mars target area in a small range, we conducted a comparative study between the Mars target area and the corresponding Qaidam basin in a small range, as shown in the Figure 13.

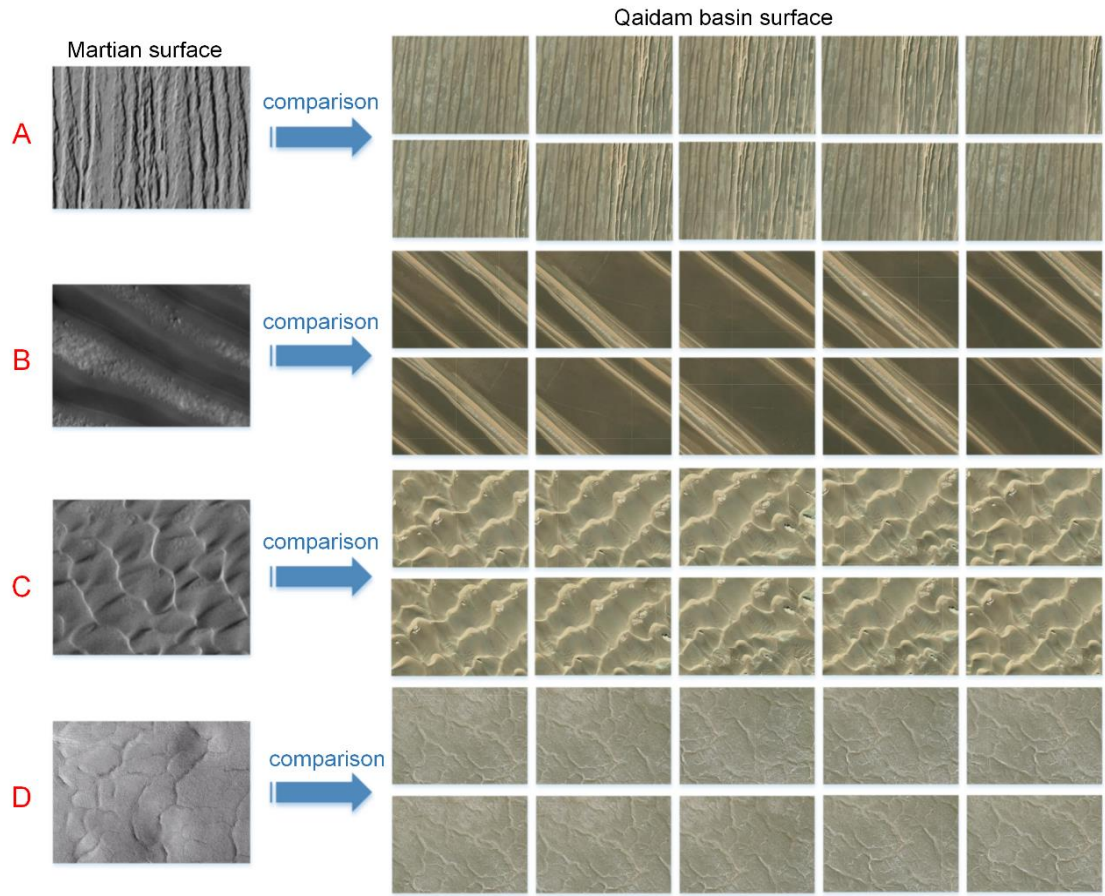

**Fig. 13** Four target regions of Mars correspond to the matching regions of the Qaidam Basin.

As seen in Figure 14, A-A', B-B', C-C', D-D' matching has the smallest difference of 0.11, 0.26, 0.38, 0.05, respectively. In the three-dimensional diagram, we also obtained the gap between each position of the four regions A, B, C and D and the Mars target region, where the horizontal coordinate represents the longitude and latitude, and the vertical coordinate represents the gap between each position and the Mars target region. We selected 10 small regions in each region A, B, C and D, and statistically found that 6 points in region A, B and C were in the 25%-75% confidence interval, and 5 points in region D were in the 25%-75% confidence interval. Among them, D region: the polygon structure on Mars is very similar to the polygon structure of the dry salt lake in the evaporite area of Qaidam Basin. The polygon structure in the salt crust is explained as the growth of salt crystals in the salt layer and mud layer under the dry salt lake crust causes the salt crust to split into polygons during the dry phase of the salt lake cycle (Christiansen, 1963). The evaporite region is a priority area for the search for life on Mars.

After selecting areas with polygonal structures on Mars similar to those in the evaporite area of Qaidam Basin, the mineral composition of the polygonal structure can be determined through methods such as hyperspectral and thermal emission imaging analysis (Osterloo et al., 2008; Wray et al., 2019). If they are halite and gypsum, priority can be given to landing on Mars and exploring signs of life (Conner and Benison, 2013; Benison and Karmanocky III, 2014). And it is possible to simulate the landing and sampling of a Mars rover in the polygonal structure of the evaporites area in the Qaidam region. At the same time, it can simulate the evaporite samples collected in the area of Qaidam region and conduct in-situ comprehensive spectral analysis using instruments carried by a Mars rover (Liu, 2018; Turenne et al., 2022).

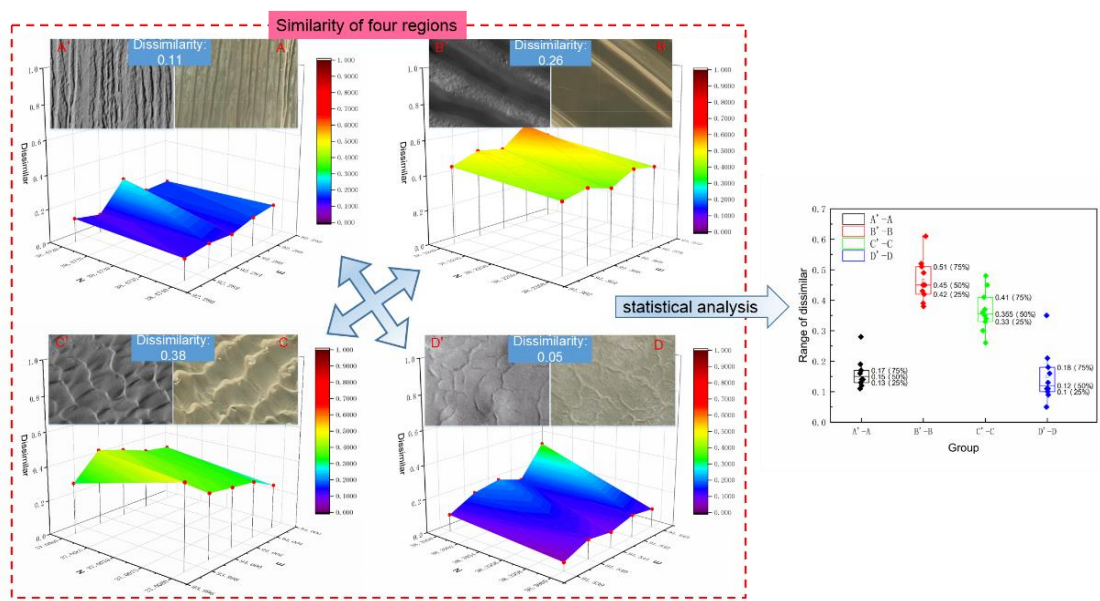

**Fig. 14** Similarity analysis of the four target regions of Mars with the corresponding Qaidam Basin probe sites.

**Discussion**

Our methodological approach successfully identified various aeolian sand dunes and evaporites from the Qaidam Basin. During the evaporation process of modern seawater/salt lake water, as salinity increases, carbonates are first deposited, followed by gypsum deposits. As salinity continues to increase, halite will be deposited, and eventually potassium salt will be deposited. Among these evaporite minerals, the largest volume is halite deposition (Warren, 2010). Therefore, if seawater is completely evaporated, the largest mineral in volume should be halite. Ice cores and halite are the best media for preserving ancient microorganisms, and the polygonal structure is the most obvious feature of the formation of salt crusts in modern salt lake evaporite deposits on a large scale (Christiansen, 1963).

Evaporite deposits are treasure troves of preserving ancient environmental and biological information, with halite being the most important mineral after the complete drying up of the evaporite basin (Meng et al., 2015, 2018 a,b). Halite minerals have good sealing properties and can solidify into halite rock in shallow burial states (70m or so), without pores and are incompressible in volume (Casas and Lowenstein, 1989). Fluid inclusions in halite, gypsum, glauberite, and mirabilite are fluids (liquids and/or gases) captured during the process of crystallization and precipitation through evaporation in the environment of ocean lagoons or land salt lakes (Lowenstein et al., 2001). Therefore, primary fluid inclusions in halite can be well preserved inside the halite minerals, recording information on the temperature, chemical composition, and atmospheric composition of the original ocean/salt lake, providing excellent direct records for the ancient environment, and preserving information on the original hydrosphere, atmosphere, and biosphere (Meng et al., 2018 a,b). Therefore, the study of evaporite regions has become a priority area for informing future Mars exploration, with a strong emphasis on polygonal structures (Zheng et al., 2013; Xiao et al., 2017).

Petrographically, fluid inclusions in halite can be divided into primary and secondary. Fluid inclusions in halite can be roughly divided into three types in petrography: 1. It has alternating light and dark bands, fluid inclusions captured during sedimentation, and can directly record the composition of the brine during evaporation; 2. Isolated large negative crystalline fluid inclusions in halite that appear at the outer edge of primary inclusion bands, formed during sedimentary or early diagenetic stages, with fluid compositions that are the same or similar to those of primary fluid inclusions during evaporation; 3. Coarse and transparent halite particles can form in different sedimentary stages after sedimentation, and the size of the inclusions varies greatly. There can be either very small inclusions or large and irregular inclusions. The composition of brine in these inclusions varies greatly, but the compositional signals of previous water bodies can still be recorded (Kovalevich et al., 1998) (Fig. 15).

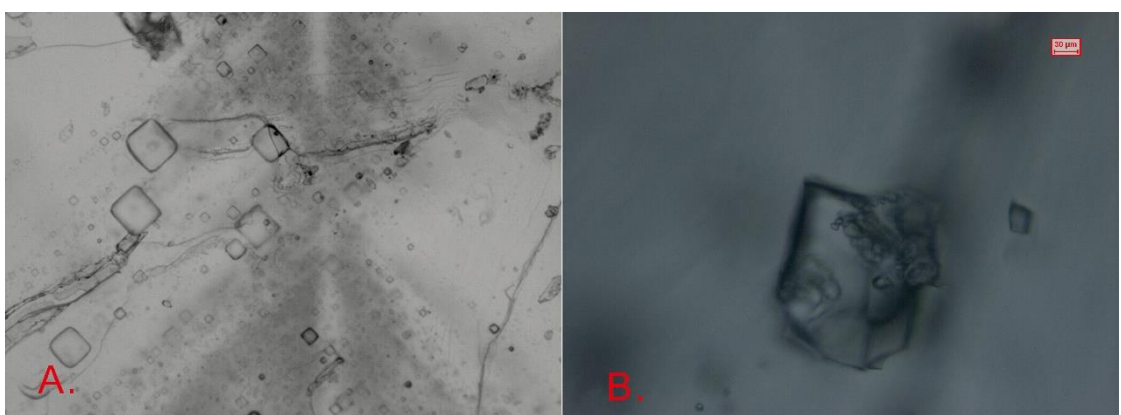

**Fig. 15** Primary and secondary fluid inclusions in halite.

The primary fluid inclusions captured in halite can be further divided into two types: i) hopper crystals, captured at the gas-brine interface, and ii) chevron crystals, grown at the bottom of the brine (Dellwig, 1955; Lowenstein and Hardie, 1985; Meng et al., 2013). Overall, primary fluid inclusions in halite can record the brine temperature of salt crystallization (i.e, the temperature of ancient seawater/salt lakes), with hopper crystals recording surface brine temperatures while chevron crystals record the bottom temperature. Hopper cystals can also capture small amounts of air as well as various algae and microorganisms in the water (Connor and Benison, 2013) (Fig.16).

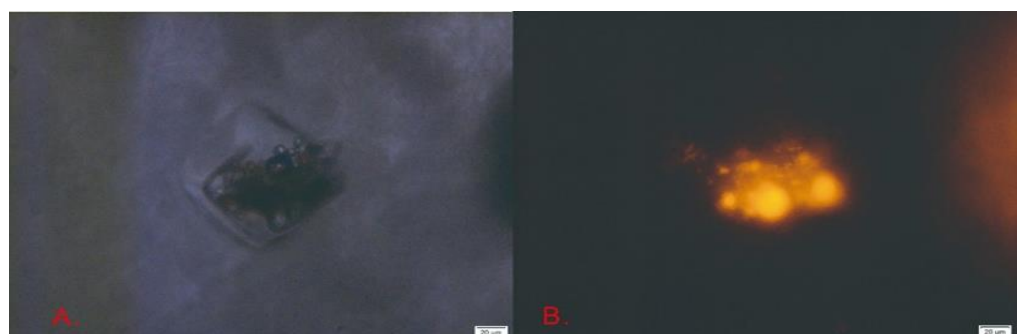

**Fig. 16** Organic matter preserved in primary fluid inclusions of Eocene halite collected in eastern China.

## Conclusion

For facilitating future Mars exploration, deep-learning approaches can be used to select the best suited areas. For astrobiology, focusing on similar landforms to the ones found in the evaporite areas of Qaidam Basin, especially those with polygonal structures, should prove particularly fruitful. After this selection, the existence of evaporite minerals at these sites on Mars can be confirmed through e.g., remote sensing technology and hyperspectral technology. Finally, a lander/rover mission can retrieve rock salt and gypsum samples, cut them along the cleavage plane, and search for signs of life from fluid inclusions.

Captions:

**Fig.1** Changes in abundance of evaporites through Phanerozoic time (revised from Ronov et al., 1980)

**Fig.2** Mineral precipitation sequence in a concentrating marine brine before bittern salts (revise from Briggs 1958)

**Fig.3** Evaporites in Qadam Basin (A. Halite; B. Mirabilite)

**Fig.4** Primary Fluid inclusions in halite from Qadam Basin

**Fig.5** A. Yadan landform similar to the surface of Mars; B. Large polygonal structures in salt crusts

**Fig.6** Process of Selecting Small-scale Martian Land Simulation Sites in the Qaidam Basin

**Fig.7** Structure of the Siamese Network Model

**Fig.8** Selection of the central region on the Martian surface and establishment of the dataset

**Fig.9** Selection of representative regions in the Qaidam Basin and establishment of the dataset

**Fig.10** Selection of local regions and data augmentation

**Fig.11** (a) Architecture of the GW function (Convolutional Network) for learning to map low-dimensional data features (b) Image similarity analysis framework based on the Siamese network model

**Fig.12** Similar matching results between the central region of the Qaidam Basin and the target regions on Mars

**Fig.13** Four target regions of Mars with four small analog points of similarity to the Qaidam Basin

**Fig.14** Similarity analysis of the four target regions of Mars with the corresponding Qaidam Basin probe sites

**Fig.15** Primary and secondary fluid inclusions in halite

**Fig.16** Organic matter preserved in primary fluid inclusions of Eocene halite of eastern China

References：

Anglés, A., Li, Y., 2017. The western Qaidam Basin as a potential Martian environmental analogue: An overview. J. Geophys. Res. Planets 122, 856-888.\

Antunes, A., Simões, M. F., Grötzinger, S. W., Eppinger, J., Bragança, J., & Bajic, V. B. (2017). Bioprospecting archaea: focus on extreme halophiles. *Bioprospecting: success, potential and constraints*, 81-112.

Baxter, B. K., Eddington, B., Riddle, M. R., Webster, T. N., and Avery, B. (2007). Great Salt

Lake halophilic microorganisms as models for astrobiology: Evidence for desiccation tolerance and ultraviolet irradiation resistance. *Instrum. Methods Missions Astrobiol.* 6694:669415. doi: 10.1117/12.732621

Bell, S., Bala, K., 2015. Learning visual similarity for product design with convolutional neural networks. ACM trans. Graph. 34, 1-10.

Benison, K., C., Karmanocky III, F. J., 2014. Could microorganisms be preserved in Mars gypsum? Insights from terrestrial examples. Geology 42, 615-618.

Bridges, J.C., Schwenzer, S.P., 2012. The nakhlite hydrothermal brine on Mars. Earth Planet. Sci. Lett. 359-360, 117-123.

Briggs, L,I,, 1958. Evaporite facies. J Sed Petrol 28:46-56

Bromley, J., Guyon, I., LeCun, Y., Sackinger, E., Shah. R., 1993. Signature verification using a siamese time delay neural network. In: Cowan, J. & Tesauro, G. (eds), Advances in neural information processing systems (NIPS 1993) 6, Morgan Kaufmann.

Burnham A. D. and Berry. A. J., 2017. Formation of Hadean granites by melting of igneous crust. Nat. Geosci. 10, 457-462.

Casas, E., Lowenstein, T.K., 1989. Diagenesis of saline pan halite: comparison of petrographic features of modern, Quaternary and Permian halites. J. Sed. Petrol. 59, 724-739.

Chopra, S., Hadsell, R. and LeCun, Y., 2005. Learning a similarity metric discriminatively with applications to face verificaton. In Proceedings of the IEEE Conference on Computer Vision and Pattern Recognition (CVPR-05) 1, 539-546.

Christiansen, F., 1963. Polygonal fracture and fold systems in the salt crust, Great Salt Lake Desert, Utah. Science 139, 607-609.

Chu, L., Li, H., Yang Z., 2020.Accurate scale estimation for visual tracking with significant deformation. IET Comput. Vis. 14, 278-287.

Conner, A. J., Benison, K.C., 2013. Acidophilic halophilic microorganisms in fluid inclusions in halite from Lake Magic, Western Australia. Astrobiology 13, 850-860.

DasSarma, S. (2006). Extreme halophiles are models for astrobiology. *Microbe Mag.* 1, 120–126. doi: 10.1128/microbe.1.120.1

Dellwig, I.F., 1955. Origin of the salina salt of Michigan. J, Sediment. Petrol. 25, 83-110.

Di Achille, G., Hynek, B. M., 2010. Ancient ocean on Mars supported by global distribution of deltas and valleys. Nat. Geosci. 3, 459-463.

Edwards, C.S., Bandfield, J.L., Christensen, P.R., Fergason, R.L., 2009. Global distribution of bedrock exposures on Mars using THEMIS high-resolution thermal inertia. J. Geophys. Res. 114, E11001.

Ehlmann, B.L., Edwards C.S., 2014. Mineralogy of the Martian Surface. Annu. Rev. Earth Planet. Sci. 42, 291-315.

El Maarry, M.R., Markiewicz, W.J., Mellon, M.T., Goetz, W., Dohm, J. M., Pack, A., 2010. Crater floor polygons: desiccation patterns of ancient lakes on Mars? J. Geophys. Res. Planets 115, E10006.

Hadsell, R., Chopra, S., LeCun, Y., 2006. Dimensionality Reduction by Learning an Invariant Mapping, 2006 IEEE Computer Society Conference on Computer Vision and Pattern Recognition (CVPR'06), New York, NY, USA, 2006, 1735-1742.

He, K., Zhang, X., Ren, S. and Sun, J. 2015. Spatial Pyramid Pooling in Deep Convolutional Networks for Visual Recognition, in: IEEE Transactions on Pattern Analysis and Machine

Intelligence, 37, 1904-1916.

Holland, H.D., 1978. The chemistry of the atmosphere and oceans. Wiley, New York.

Hui, F., Payeur, P., Cretua, A., 2017. Visual tracking of deformation and classification of non-rigid objects with robot hand probing. Robotics 6, 5.

Knauth, L.P., 2005. Temperature and salinity history of the Precambrian ocean: implications for the course of microbial evolution. Palaeogeogr. Palaeoclimatol. Palaeoecol. 219, 53-69.

Kong, F., Ma, N., 2010. Isolation and identification of halophiles from evaporates in Dalangtan Salt Lake. Acta Geol. Sin. 84, 1661-1667.

Kong, W.G., Zheng, M.P., Kong, F.J., Wang, A.L., Hu, B., 2013. Sedimentary salts at Dalangtan Playa and its implication for the formation and preservation of martian salts. Lunar and Planetary Science Conference. 1336.

Kong, W., Zheng, M., Kong, F., Chen, W., 2014. Sulfate-bearing deposits at Dalangtan Playa and their implication for the formation and preservation of martian salts. Am. Mineral. 99, 283-290.

Kovalevich, V.M., Peryt, T.M., Petrichenko, O.I., 1998. Secular variation in seawater chemistry during the Phanerozoic as indicated by brine inclusions in halite. J. Geol. 106, 695-712.

Lecun, Y., Bengio, Y., Hinton, G., 2015. Deep learning. Nature 521, 436-444.

Li, K., He, F.Z., Yu, H.P., 2018. Robust visual tracking based on convolutional features with illumination and occlusion handing. J. Comput. Sci. Technol. 33, 223-236.

Liang, M, Hu, X.L., 2015. Recurrent convolutional neural network for object recognition," 2015 IEEE Conference on Computer Vision and Pattern Recognition (CVPR), Boston, MA, 3367-3375

Liu, S., Liu, G., Zhou, H., 2019. A robust parallel object tracking method for illumination variations. Mob. Netw. Appl. 24, 5-17.

Liu, Y., 2018. Raman, Mid-IR, and NIR spectroscopic study of calcium sulfates and mapping gypsum abundances in Columbus crater, Mars. Planet. Space Sci. 163, 35-41.

Lowenstein, T.K., Hardie, L.A., 1985. Criteria for the recognition of salt-pan evaporates. Sedimentology 32, 627-644.

Lowenstein, T.K., Timofeeff, M.N., Brennan, S.T., Hardie, L.A., Demicco, R.V., 2001. Oscillations in Phanerozoic seawater chemistry: evidence from fluid inclusions. Science 294, 1086-1088.

Mayer, D., Arvidson, R., Wang, A., Sobron, P., Zheng, M., 2009. Mapping minerals at a potential Mars analog site on the Tibetan Plateau. Lunar and Planetary Science Conference, 1877.

Meng, F.W., Ni, P., Schiffbauer, J.D., Yuan, X., Zhou, C., Wang, Y., 2011. Ediacaran seawater temperature: evidence from inclusions of Sinian halite. Precam. Res. 184(1-4), 63-69.

Meng, F.W., Ni, P., Yuan, X.L., Zhou, C.M., Yang, C.H., Li, Y.P., 2013. Choosing the best ancient analogue for projected future temperatures: A case using data from fluid inclusions of middle-late Eocene halites. J. Asian Earth Sci. 67-68, 46-50.

Meng, F.W., Wang, X.Q., Ni, P., Kletetsvhka, G., Yang, C.H., Li, Y.P., Yang, Y.H., 2015. A newly isolated haloalkaliphilic bacterium from middle-lake Eocene halite formed in salt lakes in China. Carbonates Evaportes 30 (3), 321-330.

Meng, F.W., Zhang, Y.S., Galamay, A.R., Bukowski, K., Ni, P., Xing, E.Y., Ji, L.M., 2018a. Ordovician seawater composition: evidence from fluid inclusions in halite. Geol. Q. 62, 344-352.

Meng, F.W.，Zhang ZL，Zhuo QG，Ni P., 2018b. Direct Geological Records of Ancient Environments in the Evaporite Basin: Evidences from Fluid Inclusions in Halite. Bull. Mineral.

Petrol. Geochemistry 37, 451-460 (in Chinese with English abstract).

Murchie, S.L., Mustard, J.F., Ehlmann, B.L., Milliken, R.E., Bishop, J.L., McKeown, N.K., Noe Dobrea, E.Z., Seelos, F.P., Buczkowski, D.L., Wiseman, S.M., Arvidson, R.E., Wray, J.J., Swayze, G., Clark, R.N., Des Marais, D.J., McEwen, A.S., Bibring, J.P., 2009. A synthesis of Martian aqueous mineralogy after 1 Mars year of observations from the Mars Reconnaissance Orbiter. J. Geophys. Res. 114, E00D06.

Nam, H., and Han, B., 2016. Learning Multi-domain Convolutional Neural Networks for Visual Tracking," 2016 IEEE Conference on Computer Vision and Pattern Recognition (CVPR), Las Vegas, NV, USA, 4293-4302.

Ning, J,, Zhang, L., Zhang, D., Wu, C., 2012. Scale and orientation adaptive mean shift tracking. IET Comput. Vis. 6, 52-61.

Osterloo, M. M., Hamilton, V. E., Bandfield, J. L., Glotch, T. D., Baldridge, A. M., Christensen, P. R., Tornabene, L.L., Anderson, F.S., 2008. Chloride-Bearing Materials in the Southern Highlands of Mars. Science 319, 1651-1654.

Press W.H., Flannery B.P., Teukolsky S.A., Vetterling W.T., 1990. Numerical Recipes: The Art of Scientific Computing, Cambridge University Press, Cambridge, New York, New Rochelle, Melbourne, Sydney, xxii + 759 pp.

Ronov, A.B., Khain, V.E., Balukhovsky, A.N., Seslavinsky, K.B., 1980. Quantitative analysis of Phanerozoic sedimentation. Sedimentary Geology, 25: 311–325.

Saleh, K., Szénási, S., Vámossy, Z., 2021. Occlusion handling in generic object detection: A review. 2021 IEEE 19th World Symposium on Applied Machine Intelligence and Informatics (SAMI).

Scheller, E. L., Ehlmann, B. L., Hu, R.Y., Adams, D. J., Yung, Y.L., 2021. Long-term drying of Mars by sequestration of ocean-scale volumes of water in the crust. Science 372, 56-62.

Simonyan, K., Zisserman, A., 2014. Very deep convolutional networks for large-scale image recognition. arXiv preprint arXiv:1409.1556.

Stan-Lotter, H., and Fendrihan, S. (2015). Halophilic *Archaea*: Life with desiccation, radiation and oligotrophy over geological times. *Life* 5, 1487–1496. doi: 10.3390/life5031487

Tosca, N.J., McLennan, S.M., 2006. Chemical divides and evaporite assemblages on Mars. Earth Planet. Sci. Lett. 241, 21-31.

Turenne, N., Parkinson, A., Applin, D.M., Mann, P., Cloutis, E.A., Mertzman, S.A., 2022. Spectral reflectance properties of minerals exposed to martian surface conditions: Implications for spectroscopy-based mineral detection on Mars. Planetary Space Sci. 210. 105377.

Wang, J., Yang, Y., Mao, J., Huang, Z., Huang, C. and Xu, W., 2016. CNN-RNN: A Unified Framework for Multi-label Image Classification. 2016 IEEE Conference on Computer Vision and Pattern Recognition (CVPR), Las Vegas, NV, USA, 2285-2294.

Wang, N., Dai, F., Liu, F.X., Zhang, G.M., 2018. Dynamic Obstacle Avoidance Planning Algorithm for UAV Based on Dubins Path. Algorithms and Architectures for Parallel Processing, 11335, 367-377.

Warren, J.K., 2010. Evaporites through time: tectonic, climatic and eustatic controls in marine and nonmarine deposits. Earth-Sci. Rev. 98, 217-268.

Wray, J.J., 2019. Diverse Surface Mineralogy of Mars from Hyperspectral Sensing. IGARSS 2019 IEEE International Geoscience and Remote Sensing Symposium, Yokohama, Japan, 4908-4910

Wu, J. H., McGenity, T. J., Rettberg, P., Simões, M. F., Li, W. J., & Antunes, A. (2022). The archaeal class Halobacteria and astrobiology: Knowledge gaps and research opportunities. *Frontiers in Microbiology*, *13*, 1023625.

Xiao, L., Wang, J., Dang, Y., Cheng, Z., Huang, T., Zhao, J., Xu, Y., Huang, J., Xiao, Z. and Komatsu, G., 2017. A new terrestrial analogue site for Mars research: The Qaidam Basin, Tibetan Plateau (NW China), Earth Sci. Rev., 164, 84-101.

Yan, J. Jiu, B., Liu, H., Chen B. and Bao, Z., 2015. Prior Knowledge-Based Simultaneous Multibeam Power Allocation Algorithm for Cognitive Multiple Targets Tracking in Clutter. In: IEEE Transactions on Signal Processing 63, 512-527.

Yin, A., Dang, Y.Q., Zhang, M., Chen, X.H., McRivette, M.W., 2008b. Cenozoic tectonic evolution of the Qaidam basin and its surrounding regions (part 3): structural geology, sedimentation, and regional tectonic reconstruction. Geol. Soc. Am. Bull. 120 (7-8), 847-876.

Yin, A., Dang, Y.Q., Wang,L.C., Jiang,W.M.; Zhou,S.P., Chen, X.H., Gehrels, G.E., McRivette, M.W., 2008a Cenozoic tectonic evolution of Qaidam basin and its surrounding regions (part 1): the southern Qilian Shan-Nan Shan thrust belt and northern Qaidam basin. Geol. Soc. Am. Bull. 120 (7-8), 813-846.

Zagoruyko, S., Komodakis, N., 2015. Learning to compare image patches via convolutional neural networks. Proceedings of the IEEE conference on computer vision and pattern recognition, 4353-4361.

Zheng, M., Kong, W., Zhang, X., Chen, W., Kong, F., 2013. A comparative analysis of evaporate sediments on Earth and Mars: implications for the climate change on Mars. Acta Geol. Sin. (Engl. Ed.) 87, 885-897.